\documentstyle[12pt,dina4p,epsfig]{article}
\title{
  \vskip-2cm
  {\baselineskip16pt
    \centerline{\normalsize \tt DESY 95-172 \hfill ISSN 0418-9833}
    \centerline{\normalsize \tt hep-ph/9509321 \hfill}
    \centerline{\normalsize \tt September 1995 \hfill}
  }
  \vskip2cm
  {\bf
    Inclusive Two--Jet Cross Sections in $\gamma\gamma$--Processes
    at $e^+e^-$ Colliders
  }
  \author{
    {T.\ Kleinwort,  G.\ Kramer} \\
    {II. Institut f\"ur Theoretische Physik}\thanks
     {Supported by Bundesministerium f\"ur Forschung und
     Technologie, Bonn, Germany under Contract 05\,6HH93P(5) and
     EEC Program ``Human Capital and Mobility'' through Network
     ``Physics at High Energy Colliders'' under Contract
     CHRX--CT93--0357 (DG12 COMA) }
    \\
    {Universit\"at Hamburg} \\
    {D - 22761 Hamburg, Germany}
  }
  \date{}
}
\begin{document}
\maketitle
\vspace{3cm}
\begin{abstract}
\thispagestyle{empty}
We have calculated inclusive one-- and two--jet production in
photon--photon collisions in next--to--leading order superimposing
direct, single resolved and double resolved cross sections.
The results are compared with recent experimental data from the
TOPAZ and AMY collaborations at TRISTAN. Good agreement is found
between experiment and the theoretical results.
\end{abstract}
\newpage
\setcounter{page}{1}
\section{Introduction}
Phenomena involving high transverse momenta $(p_T)$ in $\gamma$--$\gamma$
collisions with the incoming photons being real or quasi--real have been
studied for many years now, both experimentally and theoretically.
Emphasis has been on $\gamma$--$\gamma$ collisions at relatively high center
of mass energies in order to obtain information beyond the low $p_T$ region
which is determined by soft physics. This has been achieved recently at
the TRISTAN \cite{c1,c2} and LEP \cite{c3} storage rings. Two TRISTAN
collaborations, TOPAZ \cite{c1} and AMY \cite{c2} have presented data for
inclusive one--jet and two--jet cross sections with $p_T$'s up to 8 GeV.
This is the domain of perturbative QCD calculations which are expected
to be valid when there is a large energy scale in the interaction.
Then the lowest order diagram is that of the quark parton model (QPM)
(see Fig.~1a) also referred to as the direct contribution in leading
order (LO) QCD.
Here the two photons couple directly to the charge of the bare quark with
no further interactions involved, so that the final state consists of just
two jets.
In next--to--leading order (NLO), i.e. $O(\alpha^2\alpha_s)$, there exists
a contribution as, for example,
 the one in Fig.~1b where a gluon is emitted from
the internal quark line. This contribution becomes singular when the
momentum of the outgoing quark (antiquark) is collinear to the momentum of the
upper (lower) photon. This divergent contribution is dealt with by
introducing the scale dependent photon structure function at the upper
or lower vertex. This leads to the so--called `single resolved' (SR)
contribution depicted in Fig.~1c in LO, being $O(\alpha\alpha_s)$, when
 the photon structure function is factored out. NLO corrections to the
single resolved cross section, for example, produce contributions with a gluon
emitted from the internal quark line in Fig.~1d which becomes singular
in the collinear limit. These singular pieces are absorbed as before
in the scale dependent photon structure function now at the lower (upper)
vertex. This is the 'double resolved' contribution (DR)
with photon structure functions at the upper and lower vertex, respectively,
as shown in Fig.~1e.
In addition to the two high $p_T$ jets the single resolved (double resolved)
contributions have one (two) low $p_T$ ``beam jets'' or photon remnant jets.
This was first discussed in the paper by Brodsky et al. \cite{c4}
who showed that
resolved photon cross sections are of the same order as from LO direct
contributions, i.e. the QPM process. This work has recently be extended
by Drees and Godbole \cite{c5}.

In all these processes, direct (D), single resolved
(SR) and double resolved (DR) there is a hard subprocess in which two
partons scatter at high $p_T$ which can be calculated in perturbative
QCD.
The subprocess cross sections are convoluted with luminosity functions that
give the distributions of the virtual photons in an electron and one or two
photon structure functions that give the distribution of quarks and gluons
inside a photon.
The topological structure of these cross sections is equi\-valent to cross
sections for jet production in $\gamma p$ collisions. There the direct
(resolved) process corresponds to the SR (DR) component in
$\gamma\gamma$ collisions. The proton structure function must be replaced
by the photon structure function.

In LO calculations the three cross sections,
D, SR and DR are superimposed, and rates for topologies with two high $p_T$
jets plus one or two photon remnant jets can be predicted.
It is well known that LO calculations give only qualitative results. In this
order the predictions depend highly on the factorization scale, in particular
for the DR contribution. Furthermore LO results do not depend
on any jet definition which
is present in NLO and in the experimental data. Therefore all three processes
must be treated in NLO, i.e.~with higher order corrections included both in
the photon structure function and the hard subprocesses in a consistent way.
Such calculations were done by Aurenche et al.~for the inclusive single--jet
cross section at $\sqrt{S} = 58$ GeV \cite{c6}.
They compared their results with the TOPAZ data \cite{c1}
and found good argeement up to the highest $p_T$ by slightly
adjusting the non--perturbative input of the photon structure function
still consistent with the data on $F_2^\gamma$.
Besides the inclusive one--jet data from TOPAZ \cite{c1} and AMY \cite{c2}
there exist also measurements
of the inclusive two--jet cross sections by the TOPAZ \cite{c1}
and the AMY \cite{c2}
collaborations. These data have been compared so far only to LO calculations
by the authors assuming different sets of photon structure functions.
In this work we make an effort to predict these two--jet cross sections in
NLO assuming the same kinematical constraints as in the two TRISTAN
experiments. Unfortunately we are not in the position yet to calculate
all three components, the D, SR and DR cross sections up to NLO. We
have only the NLO formalism for the direct and the single resolved
cross section at our disposal. Since they yield, depending on $p_T$,
the major part of the two--jet cross section at 58 GeV we shall estimate
the DR cross section with a lowest order calculation including some
$k$ factor. This $k$ factor is estimated with the inclusive one--jet
cross section by comparing with the NLO prediction.

In section 2 we explain the formalism used to calculate the inclusive
two--jet cross section. Section 3 contains the comparison with the TOPAZ and
AMY data and a discussion of the results.

\section{Inclusive Jet Cross Sections}
As explained in the introduction the inclusive jet cross section at large
$p_T$ consists of three parts according to how the incoming photons
take part in the hard subprocesses.
If both photons couple to the quarks in the hard scattering one defines
the direct cross section $\sigma^D$ (see Fig.~1a).
For example, the inclusive two--jet cross section has the following form:
\begin{eqnarray}
\frac{d^3\sigma^D}{dp_Td\eta_1d\eta_2} & = &
\left(\frac{d^3\sigma^D(\gamma\gamma\rightarrow\mbox{jet}_1+\mbox{jet}_2)}
{dp_Td\eta_1d\eta_2}\right)_{\mbox{LO}} + \frac{\alpha_s(\mu)}{2\pi}
K^D(R,M)\label{f1}
\end{eqnarray}
In (\ref{f1}) $p_T$ denotes the transverse momentum of the
measured or trigger jet with rapidity $\eta_1$,
$\eta_2$ is the rapidity of another jet such that in the three--jet sample
these two jets have the highest and second highest $p_T$.
 The first term on
the right--hand side of (\ref{f1}) stands for the LO cross section and the
second term is the NLO correction which depends on the factorization scale
$M$, at which the initial state collinear singularity is absorbed in the
photon structure function. The variable $R$ is the usual parameter defining
the size of the jets with transverse momentum $p_T$, rapidity $\eta$ and
azimuthal angle $\phi$. When two partons fulfill the snowmass
constraint \cite{c7}
with the cone size parameter $R$ they are recombined in one jet. The same jet
definition is used in the TOPAZ and AMY analysis.

The NLO corrections $K^D(R,M)$ are calculated with the phase--space slicing
method using for the seperation of the $\gamma\gamma\rightarrow 2\mbox{ jets}$
and $\gamma\gamma\rightarrow 3\mbox{ jets}$ cross sections an invariant mass
cut--off $y$, defined as $2p_ip_j<ys$, where $s$ is the partonic center of
mass energy squared. In the direct process the cross
section for $\gamma\gamma\rightarrow q\bar{q}g$ has soft, initial and
final state collinear singularities. The $\gamma\gamma\rightarrow q\bar{q}g$
cross section is integrated over these singular regions up to the cut--off
value $y$ which isolates the respective singularities which are cancelled
against the singular contributions of the virtual corrections to
$\gamma\gamma\rightarrow q\bar{q}$ and the subtraction term at the
scale $M$ which are absorbed into the photon structure function. Outside the
cut--off region controlled by the parameter $y$ we have genuine $q\bar{q}g$
final states. In this contribution two of the partons are recombined if
they obey the snowmass constraint.
The LO contribution and the NLO correction inside the $y$ cut--off
contribute to the two--jet cross section together with the contributions
inside the cone with radius $R$. That part of the $q\bar{q}g$ contribution
not fulfilling the cone recombination condition stands for the 3--jet
cross section of which we have calculated the inclusive two--jet cross section
as a function of $p_T$, $\eta_1$ and $\eta_2$ where $p_T$ is the transverse
momentum of the measured jet or trigger jet with rapidity $\eta_1$
and $\eta_2$ is the rapidity
of another jet so that both ${p_T}_1$ and ${p_T}_2$ are the jets with
highest $p_T$. For the two--jet cross section ${p_T}_1 = {p_T}_2$.
This inclusive cross section is independent of the invariant mass parameter
$y$ used to seperate those regions of phase space that contain the soft and
collinear singularities. However the inclusive two--jet cross section
depends on the cone size $R$.

If one of the two incoming photons interacts via the quark or the gluon
component in its structure function we define
\begin{eqnarray}
&&\frac{d^3\sigma^{SR}}{dp_Td\eta_1d\eta_2}  =  \nonumber\\
&& \sum\limits_{i=q,g}\int\! dx_1
F_{i/\gamma}(x_1,M)\left(\left(\frac{d^3\sigma^{SR}
(i\gamma\rightarrow\mbox{jet}_1+\mbox{jet}_2)}{dp_Td\eta_1\eta_2}\right)_{LO}
+\frac{\alpha_s(\mu)}{2\pi}K_{i\gamma}^{SR}(R,M,\mu)\right)
\nonumber\\
&+& \sum\limits_{i=q,g}\int\! dx_2
F_{i/\gamma}(x_2,M)\left(\left(\frac{d^3\sigma^{SR}
(\gamma i\rightarrow\mbox{jet}_1+\mbox{jet}_2)}{dp_Td\eta_1\eta_2}\right)_{LO}
+\frac{\alpha_s(\mu)}{2\pi}K_{\gamma i}^{SR}(R,M,\mu)\right)
\label{f2}
\end{eqnarray}
This cross section has the same structure as the direct inclusive two--jet
cross section for photoproduction $\gamma p\rightarrow\mbox{jet}_1+
\mbox{jet}_2+X$, where the photon structure function in (\ref{f2}) is replaced
by the proton structure function. This cross section has been calculated
recently with the phase space slicing method with invariant mass cut slicing
\cite{c8}.
The developed formulas for the two--jet photoproduction cross section can be
transformed easily to the corresponding $\gamma$--$\gamma$ process. The first
term in each bracket of (\ref{f2}) is the LO cross section and
$K_{i\gamma}^{SR}$, $K_{\gamma i}^{SR}$ stand for the NLO corrections which
depend on $R$ and the two scales $\mu$ (renormalization)
and $M$ (factorization).
The cross section in (\ref{f2}) has two separate terms, depending whether
the first or the second photon interacts through its structure function
$F_{i/\gamma}$.

When both photons are resolved we have
\begin{eqnarray}
\frac{d^3\sigma^{DR}}{dp_Td\eta_1d\eta_2} & = &
 \sum\limits_{i,j=q,g}\int\! dx_1\int\! dx_2F_{i/\gamma}(x_1,M)
F_{j/\gamma}(x_2,M)\nonumber\\
&&\left(\left(\frac{d^3\sigma^{DR}(ij\rightarrow\mbox{jet}_1+\mbox{jet}_2)}
{dp_Td\eta_1d\eta_2}
\right)_{LO}
+\frac{\alpha_s(\mu)}{2\pi}K_{ij}^{DR}(R,M,\mu)\right)\label{f3}
\end{eqnarray}
This cross section has the same structure as the NLO resolved inclusive
two--jet cross section for photoproduction $\gamma p\rightarrow
\mbox{jet}_1+\mbox{jet}_2+X$, where the photon structure function at one vertex
is replaced by the proton structure function. For this cross section the
higher order corrections $K_{ij}^{DR}(R,M,\mu)$ have not been calculated yet.
However they are available for the inclusive single--jet
cross section \cite{c9} and are transformed to the $\gamma$--$\gamma$ case.
Concerning the inclusive two--jet
cross section in the double resolved case our strategy is as follows.
First we have calculated the inclusive one--jet cross section in NLO,
i.e.~with the terms $K_{ij}^{DR}$, and have compared it with the LO cross
section, so that we know the $k$ factor for the one--jet cross section.
The same $k$ factor is applied to the two--jet cross section which is
calculated in LO only. Since for $p_T\ge 4$ GeV the DR contribution is less
than 20\% this estimate of the two--jet double resolved cross section is
sufficient.

It is obvious that the NLO corrections $K^D$, $K_{i\gamma}^{SR}$ and
$K_{ij}^{DR}$ in (\ref{f1}), (\ref{f2}) and (\ref{f3}) depend on the same
kinematic variables $p_T$, $\eta_1$ and $\eta_2$ as the LO cross sections.
We specified only the dependence on the cone size $R$, on the factorization
scale $M$ and on the renormalization scale $\mu$.

Before we compare our results with the TOPAZ and AMY data we have made some
tests of the NLO corrections to the one-- and two--jet cross sections.
First we checked that these cross sections are independent of the cut--off $y$
if $y$ is chosen small enough. This was the case for $y<10^{-3}$ in all
considered cases. For $y$ above $10^{-3}$ we observed some small $y$
dependence which is caused by our approximation that we neglected contributions
$O(y)$ in the analytical contribution to the two--jet cross section.

Second we tested that, first, the sum of NLO direct and the LO single resolved
cross section and, second, the sum of the NLO single resolved cross section and
the LO double resolved cross section are independent of the factorization
scale $M$. This test was performed for the one-- and two--jet cross section
seperately under kinematic conditions as for the TOPAZ data. Similar
checks were done earlier for the photoproduction one--jet cross section
\cite{c10}.

\section{Comparison with TOPAZ and AMY Data}

In $e^+e^-$ collisions at TRISTAN the jets are produced via the exchange
of two quasi--real photons. The spectrum of these small $Q^2$ photons
is described by the Weizs\"acker--Williams approximation
\begin{eqnarray}
F_{\gamma/e}(z,E) &=&
\frac{\alpha}{\pi} \frac{(1+(1-z)^2)}{z} \ln\left(\frac{E\Theta_{\max}(1-z)}
{m_e z}\right)\label{f4}
\end{eqnarray}
where $\Theta_{\max}$ is the maximally allowed angle of the electron (positron)
which experimentally is realized by the anti--tagging counters.
$E$ is the beam energy of the electron (positron) and $z=E_{\gamma}/E$
is the fraction of electron (positron) energy transferred to the
respective photon.

In this approximation the inclusive two--jet cross section
$e^++e^-\rightarrow e^++e^-+\mbox{jet}_1+\mbox{jet}_2+X$ is obtained from
\begin{eqnarray}
&&\frac{d^3\sigma(e^++e^-\rightarrow e^++e^-+\mbox{jet}_1+\mbox{jet}_2+X)}
{dp_Td\eta_1d\eta_2} = \nonumber\\
&&\int\! dz_1 \int\! dz_2 F_{\gamma/e}(z_1,E)F_{\gamma/e}(z_2,E)\,
\frac{d^3\sigma(\gamma+\gamma\rightarrow\mbox{jet}_1+\mbox{jet}_2+X)}
{dp_Td\eta_1d\eta_2}
\label{f5}
\end{eqnarray}
where the cross section on the right--hand side stands for the interaction
between two real photons.

Depending on the experimental conditions we encounter specific values for
$\Theta_{\max}$ and eventually for the integration range of $z_1$ and $z_2$,
respectively, in (\ref{f5}).

Before presenting the results we must specify the parton distributions in
the photon, $F_{i/\gamma}$, which appear in (\ref{f2}) and (\ref{f3}).
We have chosen the NLO set of Gl\"uck, Reya and Vogt (GRV) in the
$\overline{\mbox{MS}}$ scheme \cite{c11}. In \cite{c11} the photon structure
function is given
in the so--called $\mbox{DIS}_{\gamma}$ scheme. The transformation to the
$\overline{\mbox{MS}}$ scheme is known and given in \cite{c11}.
This means that the
direct, single resolved and double resolved cross section must also be
calculated with the $\overline{\mbox{MS}}$ subtraction.
We have tested that the $DIS_{\gamma}$ scheme for the photon structure
function of GRV leads to the same results.
We choose all scales $\mu=M=p_T$ and calculate $\alpha_s$ from the
two--loop formula with $N_f=4$ massless flavours
with $\Lambda_{\overline{\mbox{{\scriptsize MS}}}}=0.2$ GeV equal to the
$\Lambda$ value of the NLO
GRV photon structure function. The charm quark is treated also as a light
flavour with the boundary condition that the charm content of the
photon structure function vanishes for $M^2 \le {m_c}^2$ ($m_c=1.5$ GeV).

The TOPAZ data are obtained for $\Theta_{\max}=3.2^\circ$ and
$z_1$, $z_2\le0.75$ \cite{c1}. The rapidities of the two jets are
restricted to $|\eta_1|$, $|\eta_2|\le0.7$ and $R=1$.
The same input is used for the calculations of the NLO cross sections. Our
result for the inclusive single jet cross sections $d\sigma(\mbox{one--jet})
/dp_T$ is shown as a function of $p_T$ in Fig.~2a.
We show four curves, the direct, single resolved, double resolved and the
sum of all three components. All three cross sections are calculated up to
NLO and the curve for the sum should be compared with the TOPAZ points.
The agreement is good at small $p_T$. At larger $p_T$, where the theory
is more trustworthy, the theoretical curve falls off stronger with increasing
$p_T$ than the data. The error bars on the data include both statistical
and systematic uncertainties. The effects of the NLO corrections are
 as follows.
If we compare with the LO result, where only the NLO corrections in the
hard scattering are removed, i.e.~the LO cross section is calculated with the
same NLO GRV photon structure function as in the NLO computation we
encounter the following $k$ factors: $k\simeq 0.9$ (direct),
$k\simeq1.1$ (single resolved), $k\simeq 1.9$ (double resolved)
and $k\simeq$ 1.0 (sum)\footnote{These are the $k$ factors for the special
cone radius $R=1$. For smaller or larger cone radii the $k$ factors would
be completely different \cite{c9}.}. We emphasize that the $k$ factor
depends on the way the LO cross section is evaluated. If we use in LO
the one--loop $\alpha_s$ with the same $\Lambda$ value the $k$ factors are:
$k\simeq 0.9$ (D), $k\simeq 0.9$ (SR), $k\simeq 1.2$ (DR)
and $k\simeq 0.9$ (sum).
In Fig.~2b we have plotted the two--jet cross
section $d\sigma(\mbox{two--jet})
/dp_T$ as a function of $p_T$. We show again the direct, single resolved,
double resolved cross section and the sum of all three contributions.
Over the whole range of $p_T$ the agreement with the
TOPAZ data is excellent. We notice that the double
resolved cross section falls off with $p_T$ somewhat stronger in relation
to the direct and single resolved components as in the one--jet cross
section. The double resolved cross section is calculated in LO with the
same $k$--factor as obtained in the inclusive one--jet cross section.
For the direct and single resolved cross section the $k$ factors
for the one-- and two--jet cross section are approximately equal, so
that we can expect this also for the double resolved cross section.

The same plots but now with the kinematical contraints of the AMY experimental
data, $\Theta_{\max}=13^\circ$, $z_1$, $z_2\le 1$ and $|\eta|\le 1$ for the
one--jet cross section and $|\eta_1|$, $|\eta_2|\le 1$ for the two--jet
cross section are shown in Fig.~3a,b. All other input is as for the TOPAZ
curves. The $k$ factors are the same as for the TOPAZ cross sections.
The agreement between theory and experiment
for the one--jet data is somewhat better than in the TOPAZ case. The low
and high $p_T$ data points agree quite well. Only in the intermediate range
$3<p_T<5$ GeV the measured cross section is smaller than the
predicted cross section. The two--jet cross section agrees very
well with the data for all $p_T$. We observe that for both experiments
the two--jet data agree somewhat
better with our prediction than the one--jet data. This shows that the NLO
cross sections correctly account for the experimental data. Also the
GRV photon structure function which gives a good fit of existing inelastic
$e\gamma$ scattering data produces the correct description of the single
resolved and double resolved cross sections.
Before stronger conclusions can be drawn from the comparison
one should investigate several
effects that could influence our predictions. Most of them would change the
cross section noticeable only at small $p_T \le 3$ GeV. Such effects are,
for example, corrections to the Weizs\"acker--Williams approximation.
Part of these corrections have been studied in \cite{c12}, including intrinsic
$p_T$ effects caused by the non--zero momentum of the virtual photons.
Other corrections come from the non--vanishing charm mass or
non--perturbative intrinsic $p_T$ effects in the photon structure functions.
All these corrections, as far as they have been estimated \cite{c6,c12},
are below the experimental errors and most of them are relevant only at
small $p_T$.

\section{Conclusions}
Differential inclusive single-- and dijet cross sections
$d\sigma/dp_T$ have been calculated in NLO for the direct, single resolved
and double resolved component as a function of $p_T$. For the double
resolved two--jet cross section the NLO corrections are estimated with
a $k$ factor taken from the inclusive one--jet cross section. The sum of
these cross sections are compared to the TOPAZ and AMY experimental data.
We obtained good agreement between measured data and the theoretical
predictions. For both experiments the two--jet data seem to agree better
with the theory, than the one--jet data which may be due to the larger
experimental errors of the two--jet data.
The single and double resolved cross sections are obtained with the
GRV photon structure function which gives also good overall agreement
with all measurements on $F_2^{\gamma}$. At large $p_T$ the direct
and the single resolved components are the most important ones,
so that only at smaller $p_T$ the photon structure function can be
constrained by the double resolved component.

\newpage
\section{Figure Caption}
\begin{enumerate}
\item Generic diagrams for jet production in $\gamma$--$\gamma$
      collisions: (a) direct production in LO, (b) direct production in NLO,
      (c) single resolved production in LO, (d) single resolved production
      in NLO and (e) double resolved production in LO.

\item $d\sigma/dp_T$ with $R=1$ as a function of $p_T$ for direct
      (dashed), single resolved (dashed--dotted) and double resolved (dotted)
      production in NLO. Full curve is the sum of all three components
      compared to TOPAZ data \cite{c1}:
 \begin{enumerate}
   \item Inclusive one--jet cross section with $\eta<0.7$,
         $\Theta_{\max}=3.2^\circ$.
   \item Inclusive one--jet cross section with $\eta_1$, $\eta_2<0.7$,
         $\Theta_{\max}=3.2^\circ$.
 \end{enumerate}
         The error bars on the data include both statistical and systematic
         uncertainties.

\item $d\sigma/dp_T$ with $R=1$ as a function of $p_T$ for direct
      (dashed), single resolved (dashed--dotted) and double resolved (dotted)
      production in NLO. Full curve is the sum of all three components
      compared to AMY data \cite{c2}:

\begin{enumerate}
   \item Inclusive one--jet cross section with $\eta<1.0$,
         $\Theta_{\max}=13^\circ$.
   \item Inclusive one--jet cross section with $\eta_1$, $\eta_2<1.0$,
         $\Theta_{\max}=13^\circ$.
 \end{enumerate}
       The error bars on the data include both statistical and systematic
       uncertainties.
\end{enumerate}
\newpage

\begin{figure}[th]
 \begin{center}
  \begin{minipage}[t]{5cm}
   \begin{picture}(5,5)
    \epsfig{file=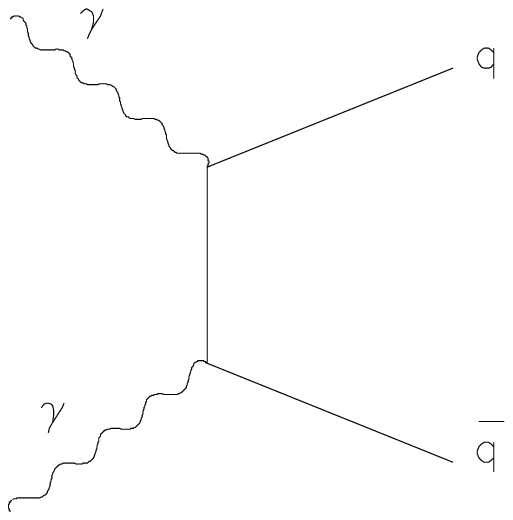,bbllx=220pt,bblly=345pt, bburx=370pt,bbury=490pt,%
            height=5cm}
   \end{picture}\par
   \begin{center} (a) \end{center}
  \end{minipage}
  \hphantom{AAAA}
  \begin{minipage}[t]{5cm}
   \begin{picture}(5,5)
    \epsfig{file=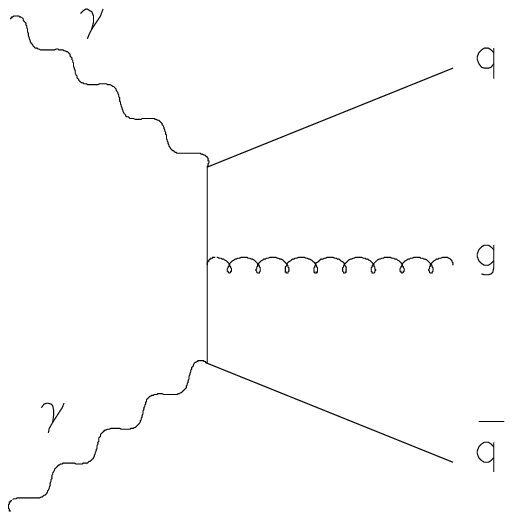,bbllx=220pt,bblly=345pt, bburx=370pt,bbury=490pt,%
            height=5cm}
   \end{picture}\par
   \begin{center} (b) \end{center}
  \end{minipage}
 \end{center}
\end{figure}

\begin{figure}[th]
 \begin{center}
  \begin{minipage}[t]{5cm}
   \begin{picture}(5,5)
    \epsfig{file=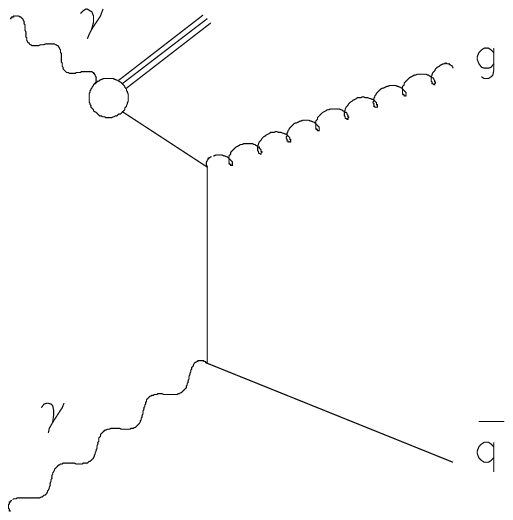,bbllx=220pt,bblly=345pt, bburx=370pt,bbury=490pt,%
            height=5cm}
   \end{picture}\par
   \begin{center} (c) \end{center}
  \end{minipage}
  \hphantom{AAAA}
  \begin{minipage}[t]{5cm}
   \begin{picture}(5,5)
    \epsfig{file=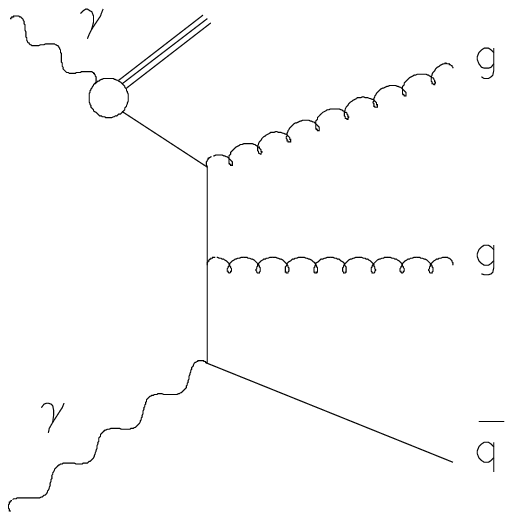,bbllx=220pt,bblly=345pt, bburx=370pt,bbury=490pt,%
           height=5cm}
   \end{picture}\par
   \begin{center} (d) \end{center}
  \end{minipage}
 \end{center}
\end{figure}

\begin{figure}[ht]
 \begin{center}
  \begin{minipage}[t]{5cm}
   \begin{picture}(5,5)
    \epsfig{file=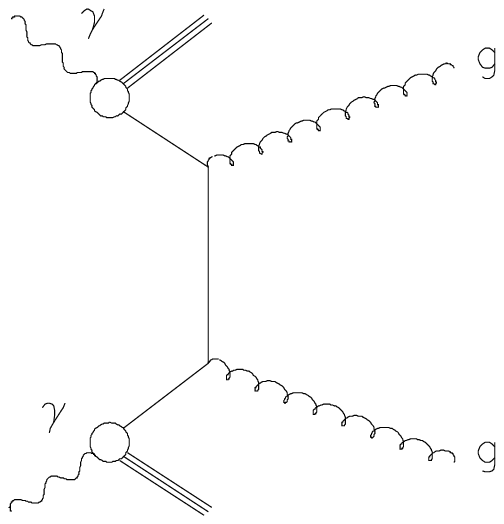,bbllx=220pt,bblly=345pt, bburx=370pt,bbury=490pt,%
            height=5cm}
   \end{picture}\par
   \begin{center} (e) \end{center}
  \end{minipage}
  \hphantom{AAAA}
  \hphantom{
   \begin{minipage}[t]{5cm}
    \begin{picture}(5,5)
     blubb
    \end{picture}
   \end{minipage}
  }
 \end{center}

\begin{center}
  Fig.~1 a-e
\end{center}
\end{figure}

\begin{figure}[t]
 \begin{center}
  \begin{picture}(15,10)
   \epsfig{file=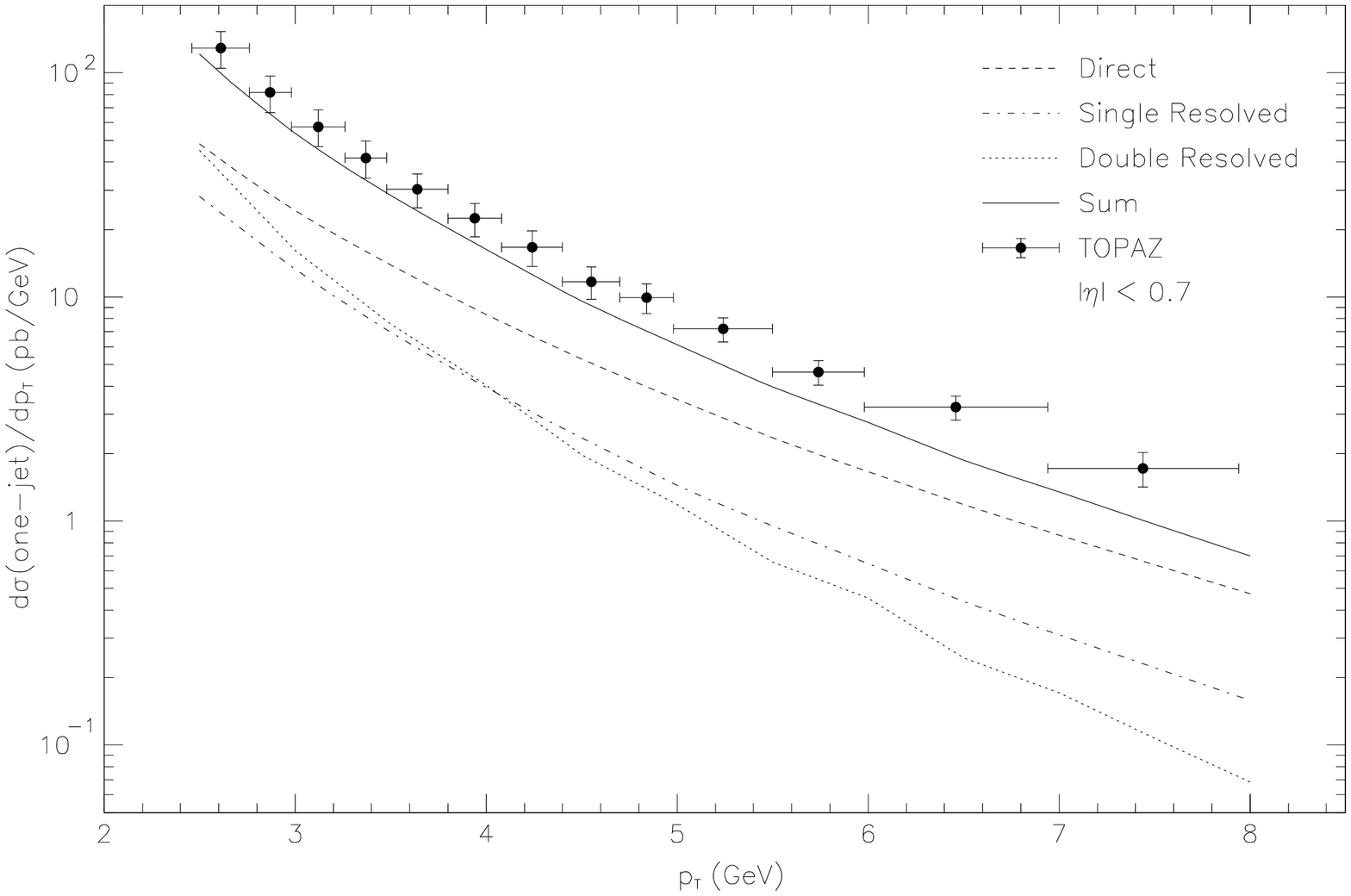,bbllx=100pt,bblly=90pt, bburx=510pt,bbury=710pt,%
           height=10cm,clip=,angle=-90}
  \end{picture}
   \begin{center} Fig.~2a \end{center}
 \end{center}

 \begin{center}
  \begin{picture}(15,10)
   \epsfig{file=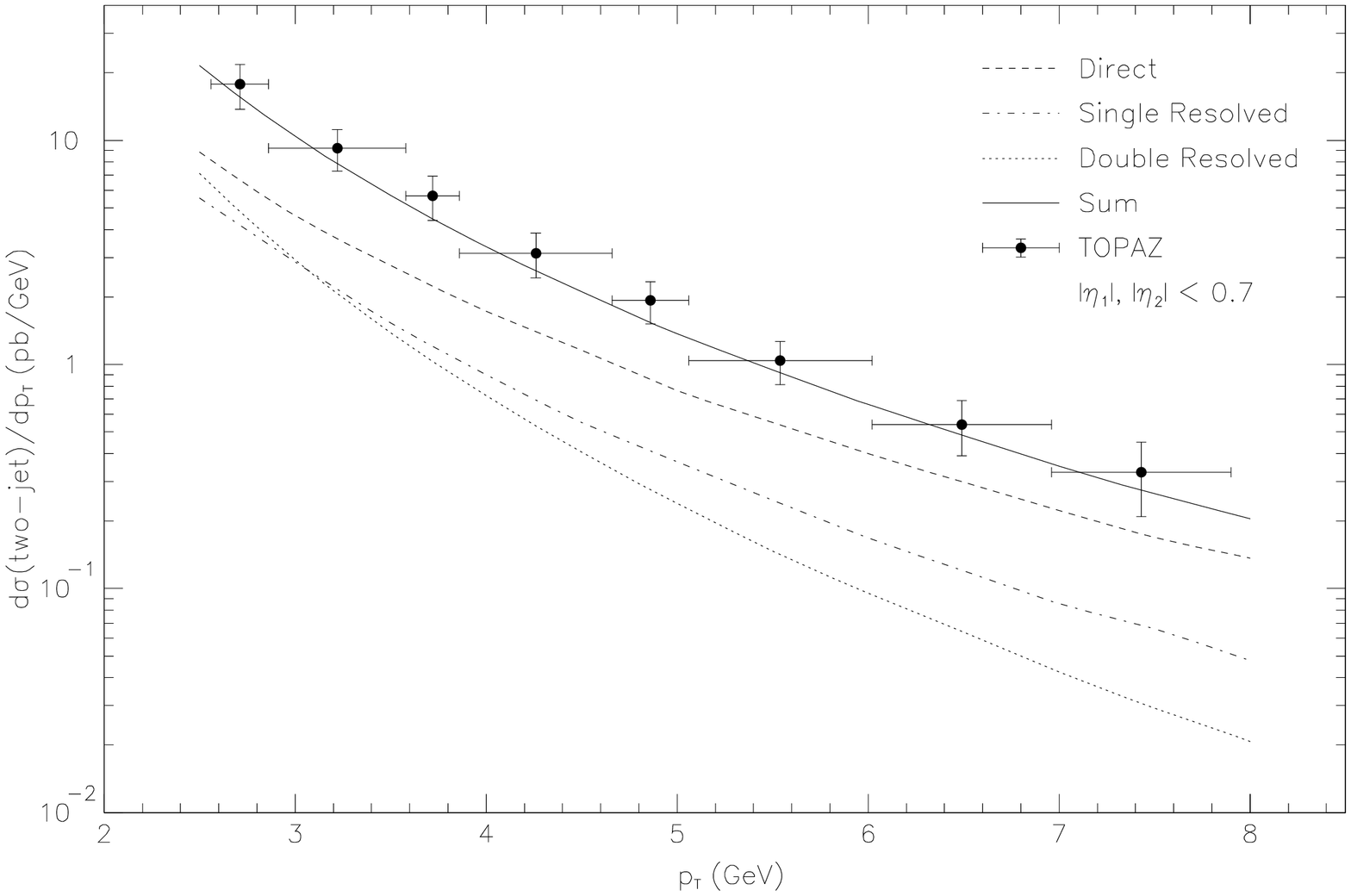,bbllx=100pt,bblly=90pt, bburx=510pt,bbury=710pt,%
           height=10cm,clip=,angle=-90}
  \end{picture}
   \begin{center} Fig.~2b \end{center}
 \end{center}

\end{figure}

\begin{figure}[t]
 \begin{center}
  \begin{picture}(15,10)
   \epsfig{file=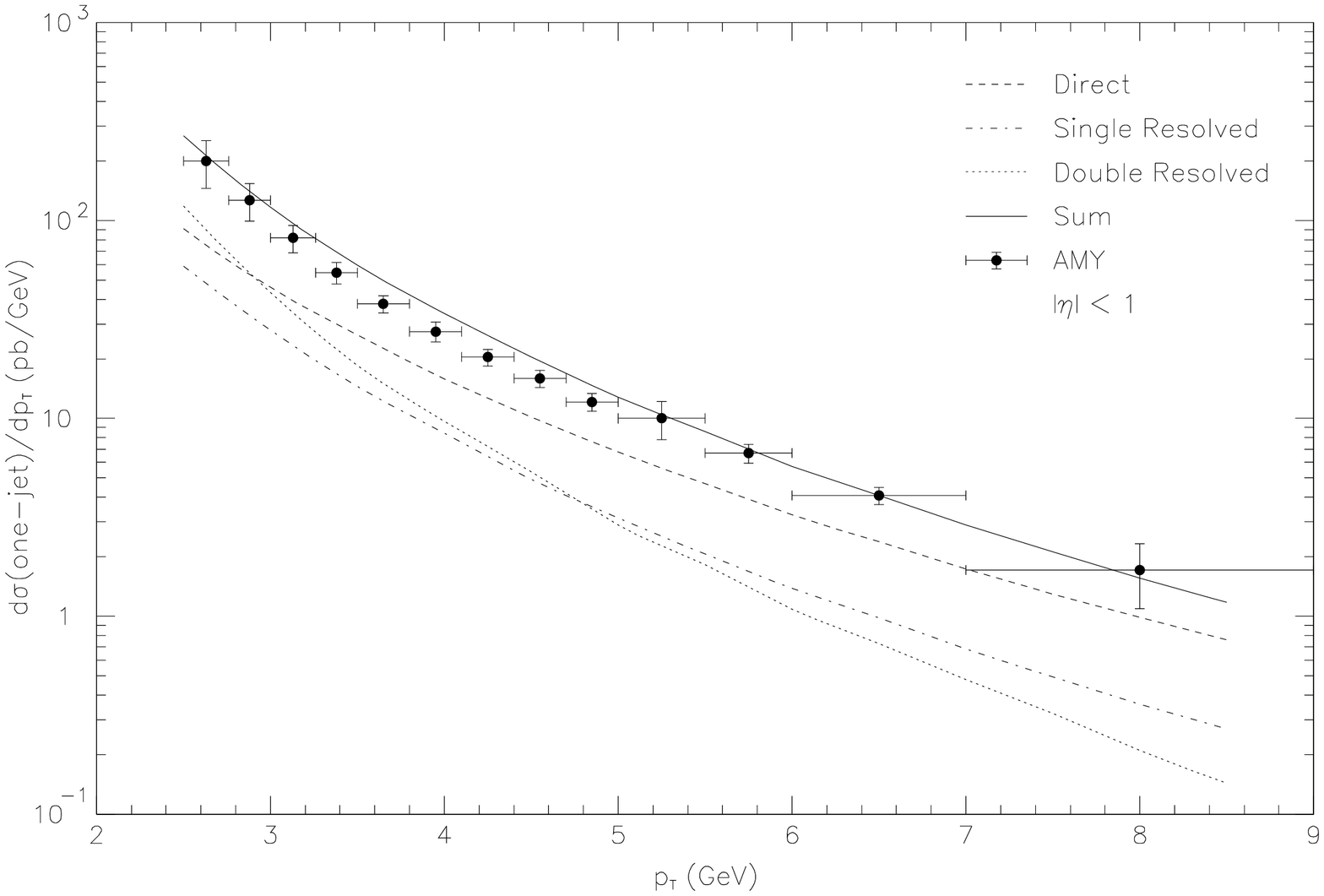,bbllx=100pt,bblly=90pt, bburx=510pt,bbury=710pt,%
           height=10cm,clip=,angle=-90}
  \end{picture}
   \begin{center} Fig.~3a \end{center}
 \end{center}

 \begin{center}
  \begin{picture}(15,10)
   \epsfig{file=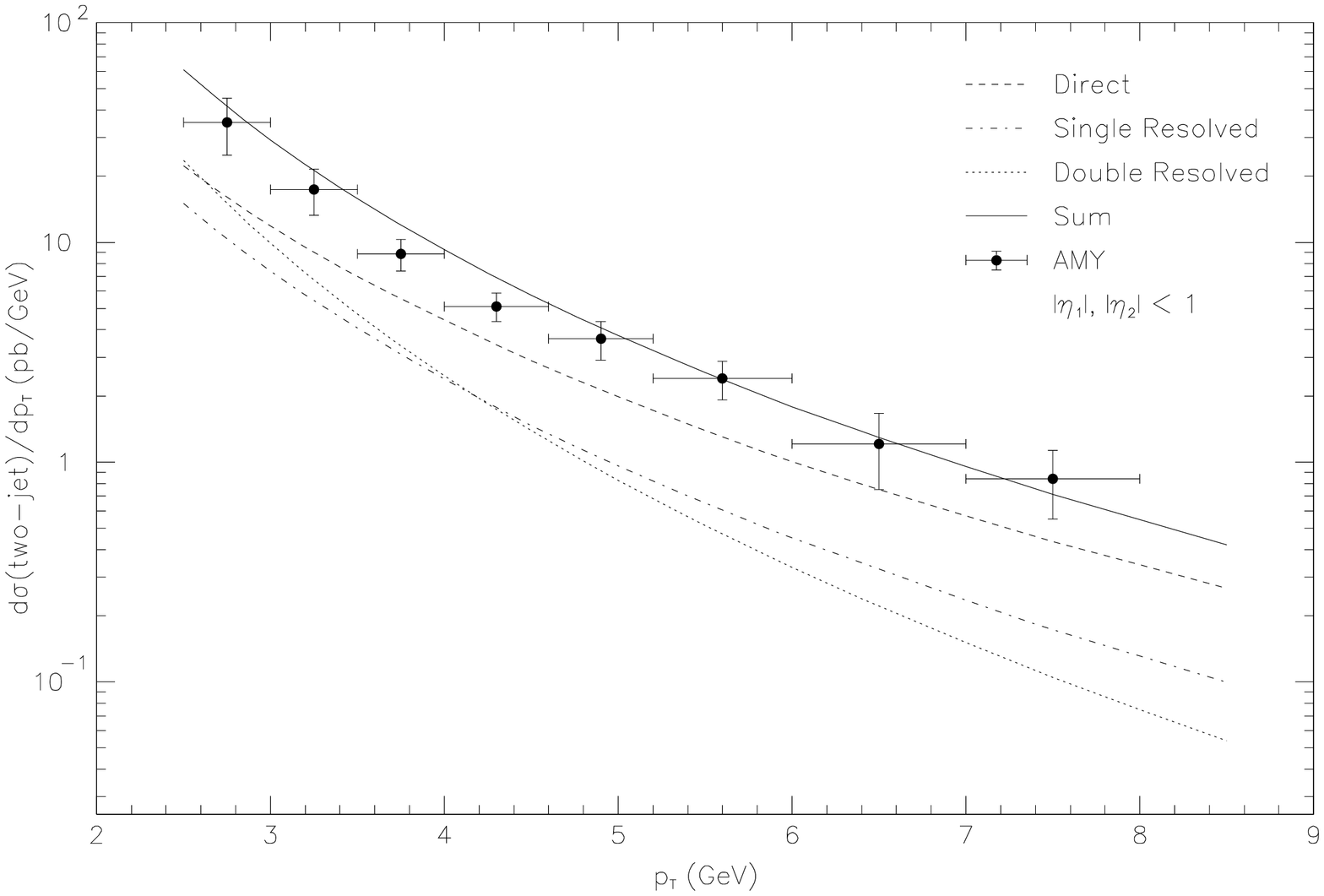,bbllx=100pt,bblly=90pt, bburx=510pt,bbury=710pt,%
           height=10cm,clip=,angle=-90}
  \end{picture}
   \begin{center} Fig.~3b \end{center}
 \end{center}

\end{figure}

\end{document}